\documentclass[prd,aps,epsfig,showpacs,twocolumn,floats,superscriptaddress]{revtex4}

\usepackage{graphicx}

\newcommand{\be}{\begin{equation}}
\newcommand{\ee}{\end{equation}}
\newcommand{\bea}{\begin{eqnarray}}
\newcommand{\eea}{\end{eqnarray}}
\newcommand{\ba}{\begin{array}}
\newcommand{\ea}{\end{array}}

\begin{document}
\title{Simple generalizations of Anti-de Sitter space-time}
\newcommand{\addressImperial}{Theoretical Physics, Blackett Laboratory, Imperial College, London, SW7 2BZ, United Kingdom}

\author{Jo\~{a}o Magueijo} 
\affiliation{\addressImperial}

\author{Ali Mozaffari} 
\affiliation{\addressImperial}

\begin{abstract}
{We consider new cosmological solutions which generalize the cosmological
patch of the Anti-de Sitter (AdS) space-time,
allowing for fluids with equations of state such that $w\neq -1$. We use 
them to derive the associated full manifolds. We find that
these solutions can all be embedded in flat five-dimensional space-time
with $--+++$ signature, revealing deformed hyperboloids. The topology 
and causal-structure of these spaces is therefore unchanged, and 
closed time-like curves are identified, before a covering space is considered. 
However the structure of Killing vector fields is entirely different
and so we may expect a different structure of 
Killing horizons in these  solutions.}
\end{abstract}
\pacs{0000000}
\maketitle

\bigskip

\section{Introduction}
Anti-de Sitter (AdS) space-time~\cite{Hawking Ellis, Gibbons2} is a 
maximally symmetric solution 
to Einstein's field equations with a negative cosmological 
constant $\Lambda$.  It is one of the simplest solutions to Einstein's
gravity and as such it has been a prime test ground for new ideas 
and toy models in (quantum) gravity. More recently, AdS is best
known for its role in the 
AdS/CFT correspondence~\cite{Maldacena,gubs,witt,Maldcena et al}, 
which conjectures that string
theories on a given space are dual to conformal field theories on the conformal
boundary of this space.  Typically the space in question
is the product of AdS with a closed manifold. For example 
type IIB string theory on AdS$_5\,\times\,$S$^5$ is dual 
to $\mathcal{N}=4$ SYM on the 4D boundary of AdS$_5$. 

A natural question is whether more realistic space-times might support
extensions of conjectures made (or theorems proved) 
for AdS. For this reason it
is interesting to consider deformations of AdS, i.e. families of
solutions to Einstein gravity which contain AdS as a limiting case
(allowing, by suitably varying a parameter, to be as close as wanted 
to AdS).

It is well known that a portion of AdS can be discovered 
using the formalism
of homogeneous and isotropic cosmology. The full manifold can then be inferred
by extension. In this paper we consider cosmological solutions that 
follow from altering the equation
of state $w=p/\rho$ (where $p$ is the pressure and $\rho$ is the energy  
density). AdS follows from $w=-1$, but a variety of solutions with
similar properties result from $w<-1/3$. 
In Section~\ref{deriv} we derive these solutions and in Section~\ref{full}
we use them to infer their associated inextendible manifolds. 
Finally in Section~\ref{disc} we carry out a preliminary study of 
the local and global properties of these solutions.

\section{Cosmological solutions}\label{deriv}
A patch of AdS may be
discovered using the formalism of Friedmann-Roberstson-Walker (FRW)
homogeneous and isotropic 
cosmology. As is well known, for a constant equation of state
$w=p/\rho$, the continuity equation,
\be \dot{\rho} + 3\,{\dot{a} \over a}\,\rho\,(1 + w) = 0 
\label{continuity eqn}\ee
integrates into $\rho \propto a^{-3(1 + w)}$ so that
the Friedmann equation: 
\be \left({\dot{a} \over a}\right)^2 = {8\pi\,G \over 3}\,\rho 
- {k \over a^2} + {\Lambda \over 3}
\label{friedmann eqn}\ee 
becomes
\be \dot{a}^2 = \frac{\Lambda a^2}{3} + C \,a^{\beta} - k 
\label{feqn beta}
\end{equation} 
where $\beta = - (1 + 3w)$ and $C$ is a constant.
The cosmological portion of AdS follows 
from $k = -1$, $w=-1$, so that (\ref{feqn beta})
becomes \be\dot{a}^2 = 1 + \left({\Lambda
\over 3} + C \right)a^2\; ,\ee 
with the extra condition $C+\Lambda/3<0$. Setting 
$\omega^2=-(C+\Lambda/3)$ converts (\ref{friedmann eqn})
into a simple harmonic oscillator equation:
\be \dot{a}^2 = 1-  \omega^2 a^{2}\label{feqn AdS} \ee  
solved by $a = \cos\omega t$.  
The FRW form of the AdS$_4$ is therefore:
\be\label{adsmetric}
ds^2 = -dt^2 + \cos^2{(\omega t)} \,d\sigma
\ee
where $d\sigma$ is the metric on a 3D homogeneous space negatively curved:
\be
d\sigma = {dr^2\over {1 + r^2}} +  r^2\,d\Omega_2 
\ee 
where $d\Omega_2=d\theta^2 + \sin^2\theta\, d\phi^2$ 
is the area element of a 2-sphere.

In this construction
we can use interchangeably a negative cosmological constant 
or a fluid with $w=-1$ and negative energy. A generalization 
can be obtained by considering a fluid with negative energy ($C<0$),
but with any $w<-1/3$ (with $\Lambda=0$ and 
$k = -1$). To fix ideas set $C=-1$ (but a generalization is 
straightforward). Then the Friedman equation becomes
\be \dot{a}^2 = 1 - a^{\beta} \label{feqn2 beta}\; ,
\ee 
where $\beta=-(1+3w)>0$.
This is no longer a simple harmonic oscillator equation but, availing ourselves of
the diffeormorphism invariance of relativity, we may define a time
variable $\eta$ via: 
\be 
dt^2 = \left(\frac{1 - a^2}{1 - a^{\beta}}\right)\, d{\eta}^2 \; .
\label{t eta relation}
\ee 
Time coordinate $\eta$ is not proper time for a cosmological observer,
but, rather, a  ``harmonic'' time for which Eq.~(\ref{feqn2 beta}) reduces to 
\be a'^2 = 1 - a^2 \label{a harmonic}\ee 
where $ a' = da / d\eta $. Thus $a = \cos \eta$ leading to metric:
\be 
ds^2 = 
-\left(\frac{\textrm{sin}^2\,\eta}{1 - \textrm{cos}^{\beta} \eta}\right)\, d{\eta}^2 + \textrm{cos}^2 \eta \,d\sigma \label{general AdS metric} \ee 
so that with these coordinates only $g_{00}$ is modified. 

For $\beta = 2$, (i.e. $w=-1$) metric  (\ref{general AdS metric}) 
reduces to the AdS metric.  For $\beta = 0$ (corresponding to $w = -1/3$, 
i.e. the Milne universe), the $g_{00}$ component becomes singular.
For other $\beta>0$ ($w<-1/3$)   
we have obtained generalizations of the 
 AdS solution.  They can only be considered in the
Friedmann patch of the manifold, and hence it would be interesting
to seek their complete manifold, just like for AdS.

For completeness, we include our metric written in terms of proper time $t$. 
We note that 
\be t = \int \frac{da}{\sqrt{1 - a^{\beta}}} =
 a\;_2F_1\left(\frac{1}{2},\frac{1}{\beta};1 + \frac{1}{\beta}; a^{\beta}\right)
\ee 
where $_{2}F_1$ is a hypergeometric function. We can thus recast 
(\ref{general AdS metric}) in terms of $t$, but the result must be 
expressed in terms of special functions and their inverses.

One may wonder what kind of matter content could lead to these models. 
Most obviously one could consider a scalar field with a suitable
potential. For $K=0$ an exponential potential leads to solutions
with constant $w$, and it's possible to obtain any $w<-1/3$ (and even
a negative energy density) by carefully
choosing the sign of the kinetic and potential energy 
terms~\cite{scaling,phantom,Phantom}. Similar solutions might be obtainable
for $K=-1$ with an appropriate choice of potential. We defer this study to 
future work, but note that one often simply postulate the equation of state 
$w$ for a fluid. 

\section{The associated full manifold}\label{full}

\begin{figure}[h]
\resizebox{0.7\columnwidth}{!}{\includegraphics{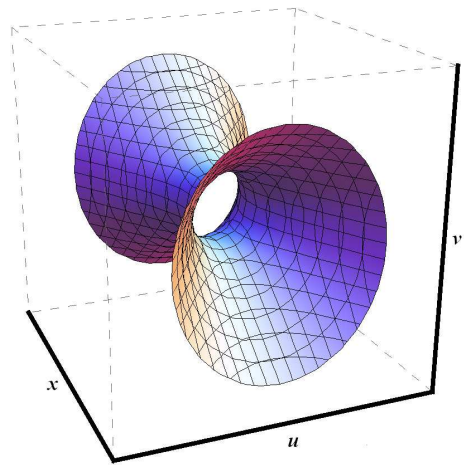}}
\caption{\label{fig:beta = 2} As is well known AdS space can be represented
by an hyperboloid living in 5D flat space with signature $--+++$.
Closed time-like curves are evident (see text; but note that any
curve in the $u,v$ plane is time-like).}  
\end{figure}

The complete extension of the above patch can be obtained 
by a simple adaptation of the  procedure for AdS, which we briefly
review here~\cite{Hawking Ellis}. Consider a 5D manifold 
 with metric 
\be \label{5Dmetric}
ds^2 = -du^2 - dv^2 + dx^2 + dy^2 + dz^2
\ee 
where there live a series of hyperboloids (see Fig.~\ref{fig:beta = 2}): 
\be u^2 + v^2 - x^2 - y^2 - z^2 = \rho^2 \; .
\ee 
We can introduce a system of coordinates $\{t,r,\rho,\theta,\phi\}$
by means of: 
\bea
u &=& \rho \sin t\nonumber\\
v &=& \rho \cos {t} \;\sqrt{1 + r^2}\nonumber\\
x &=& \rho \cos{t} \;r \cos{\theta}\nonumber \\
y &=& \rho \cos{t} \;r \sin{\theta}\cos{\phi}\nonumber\\
z &=& \rho \cos{t}  \;r \sin{\theta}\sin{\phi}
\label{AdS embeddings 1}
\eea
for which $\rho=$ constant represents the hyperboloids. 

The induced metric on the hyperboloids may be found by writing
(\ref{5Dmetric}) in terms of these coordinates and setting $d\rho=0$, an exercise that reveals the
metric (\ref{adsmetric}). Thus the portion of the 
hyperboloids covered by these coordinates are embeddings of the 
AdS Friedmann patch. For different $\rho$, different values
of $\omega$ are recovered.

To find the whole manifold, we note that there are apparent singularities 
at $t = \pm \frac{1}{2}\pi$ and therefore these coordinates do 
not cover the whole space~\cite{Hawking Ellis}. We can reframe the 
space into a system of static coordinates using 
relations~\cite{Carroll,Maldacena}:
\bea
u &=& \rho \sin{t'}\cosh{r'}\nonumber\\
v &=& \rho\cos{t'}\cosh{r'}\nonumber\\
x &=& \rho \sinh{r'}\cos{\theta}\nonumber\\
y &=& \rho \sinh{r'}\sin{\theta}\cos{\phi}\nonumber\\
z &=& \rho \sinh{r'}\sin{\theta}\sin{\phi}\; .
\label{AdS embeddings 2}
\eea
These cover the full hyperboloid and for $\rho=1$
the metric is now 
\be 
ds^2 = -\cosh^2{r'}\,dt'^2 + dr'^2 + \sinh^2{r'}\,d\Omega_2 \label{AdS full manifold} \; .\ee
By choosing different values of $\rho$ different $\Lambda$ may be implemented,
but as before, we shall consider $\rho=1$ to fix ideas in what follows.

A similar construction may be devised for the 
generalisations in Eq.~(\ref{general AdS metric}). The manifolds
they represent can be embedded in 5D flat space with the same signature.
Given that the $d\sigma$ components of the metric remain
the same we try
\bea
x &=& \rho \,a \,\sinh\chi\cos\theta\\
y &=& \rho \,a \,\sinh\chi\sin\theta\cos\phi\\
z &=& \rho \,a \,\sinh\chi\sin\theta\sin\phi 
\eea
and indeed 
$\Sigma\, dx_i^2 = \rho^2\,a^2\, 
\textrm{sinh}^2\,\chi\,d\,\Omega_2^2 + [d(\rho \, a\,\textrm{sinh}\,\chi)]^2$,
replicating that part of the AdS calculation. 
Setting $v = \rho \,a\,\cosh \chi$ and for some yet to be defined function $u_*$ we find:
\begin{eqnarray} ds^2 &=& - du_*^2 - dv^2 + dx_i^2 \nonumber\\
&=& - du_*^2 - \rho^2\,da^2 - a^2\,d\rho^2 + \rho^2\,a^2\,d\sigma\; .
\label{4-FLRW metric} 
\end{eqnarray} 
If $u_* = \rho J(\eta)$ then 
\be du_*^2 = \rho^2\,{J}'^2\,d\eta^2 + 2\rho\,{J}\,{J}'\,d\rho\,d\eta + {J}^2\,d\rho^2 \label{radius} \ee 
where ${J}' = d{J} / d\eta$. Inserting $a = \cos\eta$ and 
equating the terms in $d\eta^2$  in (\ref{general AdS metric}) 
and (\ref{4-FLRW metric}) gives: 
\begin{eqnarray}
\rho^2 {J'}^2 d\eta^2
&=& \rho^2 {\left(-\sin^2\eta + \frac{\sin^2 \eta}{1-\cos^{\beta}\eta}
\right)} d\eta^2 \nonumber \\ 
&=& \rho^2 \left ( \frac{\sin^2{\eta} 
\cos^\beta{\eta}}{{1-\cos}^{\beta}{\eta}}\right)d\eta^2\; .
\end{eqnarray}
This reduces to 
\be d J ^2 = \left ( \frac{{\sin}^2\,\eta\,{\cos}^\beta\,\eta}{{1-\cos}^{\beta}\,\eta}\right)
\, d\eta^2 \label{J equation} \ee
which can be explicitly integrated into  
\be
J(\eta) = \frac{2}{\beta + 2}\,_2\,F_1\left( \frac{1}{2} + \frac{1}{\beta}, \frac{1}{2} ; \frac{3}{2} + \frac{1}{\beta}; {\cos}^{\beta}\,\eta\right)\,{\cos}^{\,\beta/2
+ 1}\,\eta
\ee
Thus (\ref{general AdS metric}) is the metric induced on the $\rho=$const
surfaces (for which $d\rho=0$) if we take the 5D space metric to be:
\be ds^2 = - du_*^2 - dv^2 + dx^2 + dy^2 + dz^2 
\ee 
The $\rho=$ constant surfaces can be inferred by analogy, resulting in a deformed hyperboloid of form:
 \be \left(\frac{\textrm{sin}\,\eta}{J(\eta)}\right)^2
u_*^2 + v^2 - x^2 - y^2 - z^2 = \rho^2 \ee 
where $u_* = \rho J(\eta)$ \\

\begin{figure}[h] 
\resizebox{0.7\columnwidth}{!}{\includegraphics{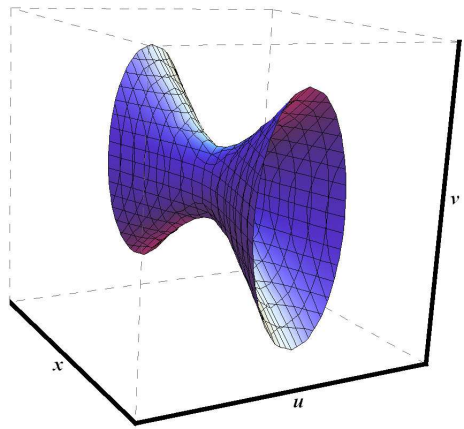}}
\caption{\label{fig:beta = 4} The full manifold corresponding to
the cosmological solution with  $w=-5/3$. For all $w<-1$
the hyperboloid squashes along the $u$ direction,
the effect becoming more pronounced the smaller the $w$.
}  
\end{figure}

\begin{figure}[h] 
\resizebox{0.7\columnwidth}{!}{\includegraphics{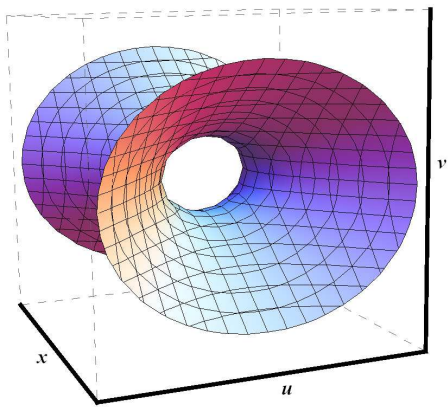}}
\caption{\label{fig:beta = 1} The full manifold corresponding to
the cosmological solution with  $w=-2/3$. For $-1>w>-1/3$ 
the hyperboloid elongates along the $u$ direction, 
the more so the closer to $w=-1/3$ (i.e. $\beta=0$) one gets.
}  
\end{figure}

We note that for $\beta = 2$ (i.e. $w=-1$) we have $J = \sin\eta$ 
and we recover the AdS construction. For $\beta = 0$, i.e. Milne space-time
(with $w=-1/3$) the construction becomes singular. 
For other values of $w<-1/3$ the result can be seen most
 effectively by plotting the resulting deformed hyperboloids.  
As Figure~\ref{fig:beta = 4} shows, if $w<-1$ the hyperboloid 
squashes in the $u$ direction, the more so the smaller the value of $w$.
For $-1<w<-1/3$, as Fig~\ref{fig:beta = 1} shows, the hyperboloids 
expand in the $u$ direction instead, the effect becoming more extreme as
$w=-1/3$ is approached.  
Since $\rho \propto a^{\beta-2}$ there are Ricci
singularities as $a \rightarrow 0$.  This means that for $a = \cos(\eta) = 0$, 
i.e. when $\eta = \pm \pi/2$, there are ``point-like'' singularities,
seen as cusps on the hyperboloid.  We pick the particular case of $\beta = 1$ 
to that find 
\be J(\eta) \propto \left(\sqrt{(1-\cos(\eta))\cos(\eta)} - \sin^{-1}(\sqrt{\cos(\eta)}\right)\ee 
which has roots at $\eta = \pm \pi/2$. The hyperboloid construction fails 
at these points (which are located on the plane $u = \rho, v=z=y=z=0$).  
These singular points on the hyperboloid are found for all $0 < \beta < 2$

\section{Discussion}\label{disc}
These manifolds may prove valuable in assessing conjectures that 
theorems proved for AdS generalize to more realistic space-times.
A close scrutiny of their properties is therefore in order.
Here we briefly discuss their most evident properties. 

From the embeddings found it's immediately obvious that the new manifolds
share with AdS its topology and aspects of the causal structure. 
In particular they all admit closed time-like curves (any curve in the
$u,v$ plane is time-like).  From the embedding of AdS we see
that the time $t'$  is periodic: $t'$ and $t' + 2\,\pi$ 
represent the same point  on the hyperboloid.  Thus any curve with 
fixed ${\rho, \theta, \phi}$ and increasing $t'$ 
is a closed time-like curve (CTC) (see Fig.~(\ref{fig:beta = 2})). 
The same feature is present for all deformed hyperboloids. By unwrapping
these circles one obtains universal covering spaces, 
like for AdS~\cite{cupbook,Hawking Ellis}.

In contrast with these AdS-like features, the structure of Killing 
vector fields is entirely changed in the new spaces.
For any 3+1 metric, there exist up to 10 Killing vector fields:
3 rotational, 3 translational, 3 boosts and 1 time-like vector. 
All 10 Killing vectors are manifest on the full AdS manifold.  
In our solutions, the rotations and translations survive, but 
the time-like and boost isometries are obviously lost.


Within these fields, a null integral hypersurface 
can sometimes be identified, known as a Killing Horizon.  Associated with these there are important geometrical and thermodynamical quantities~\cite{Townsend Blackhole}, such as the surface gravity $\kappa$, and Hawking's temperature $T = \kappa/2\pi$.
Killing horizons in AdS are highly non-trivial, however by considering optical metrics (see~\cite{Gibbons1,Gibbons2})
one finds three classes of time-like, orthogonal Killing vectors in the space.
Clearly with the loss of isometries these results do not translate into our construction, but its thermodynamical properties should be the subject of a future study.

We have been unable to find the equivalent of static coordinates for 
the new spaces, and we conjecture that they don't exist (it's not 
obvious that a simple adaptation of Birkhoff's theorem can be used
to prove this). It is easy, however, to adapt the constructions used
for building the AdS Penrose diagram, as well and inferring the key causal
features. Let us write the metric in terms of conformal time:
\be
\xi=\int d\eta \frac{|\sin{\eta}|}{\cos{\eta}\sqrt{1-\cos^\beta{\eta}}}
=\frac{2}{\beta}\tanh^{-1}[{\rm sgn}(\eta)\sqrt{1-\cos^\beta{\eta}}]
\ee
A crucial feature of AdS is that light rays may reach $\chi=\infty$
in a finite amount of affine parameter (and proper time for time-like
observers). Thus the Friedmann patch is extendable. 
This feature is also true here: $\xi(\eta)$ diverges
at $\eta=\pm \pi/2$, and although $\eta$ is not proper time, proper time is 
convergent at these points. 

\begin{figure}[h] 
\resizebox{0.75\columnwidth}{!}{\includegraphics{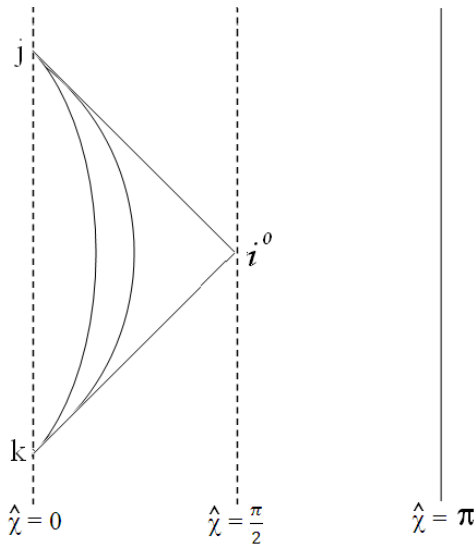}}
\caption{\label{fig:AdS Penrose} The Penrose Diagram for our space-time
when $\beta>2$. Here $k$ and $j$ 
are observers at $\hat\chi = 0$; all cosmological observers move from $k$ to
$j$ inside the half-diamond depicted (to be repeated up and down the diagram).
The diagram can be extended to the whole
Einstein static Universe, i.e. up to $\hat \chi=\pi$. Points at 
$\hat\chi=\pi/2$ and $\hat\xi=n\pi$ represent spatial infinity.}  
\end{figure}

The metric can be written in terms of $\xi$ by noting that
\be
\cos^\beta(\eta)={\rm sech}^2\frac{\beta\xi}{2}
\ee
so that:
\be
ds^2={\rm sech}^{\frac{4}{\beta}}{\left(\frac{\beta\xi}{2}\right)}
[-d\xi^2+d\chi^2
+\sinh^2\chi d\Omega_2]
\ee
We can now perform the usual conformal transformation that maps
the metric into a diamond inserted in the Einstein
Static Universe (ESU). Specifically we set up null coordinates:
\bea
u&=&\xi-\chi\\
v&=&\xi+\chi
\eea
make infinity tangible (and extendable) via:
\bea
\tan p&=&{\rm tanh}\frac{u}{2}\label{tan1}\\
\tan q&=&{\rm tanh}\frac{v}{2}\label{tan2}
\eea
and unwrap the new null coordinates into a new space and time:
\bea
p&=&\frac{\hat\xi-\hat\chi}{2}\\
q&=&\frac{\hat\xi+\hat\chi}{2}\; .
\eea
This leads to a metric conformal to the ESU: 
\be
ds^2=\Omega^2(-d\hat\xi^2 + d\hat\chi^2 +\sin^2\hat\chi d\Omega_2)
\ee
where the initial cosmological patch corresponds to the diamond
$-\pi/4<p<q<\pi/4$ (see Fig.~\ref{fig:AdS Penrose}). 
The conformal factor is: 
\be
\Omega^2={\rm sec}(2p)\, {\rm sec}(2q)\, 
{\rm sech}^{\frac{4}{\beta}}{\left(\frac{\beta\xi(p,q)} {2}\right)}\,
\ee
and for $\beta=2$ this reduces to 
$\Omega^2=\sec^2{\hat \chi}$, as is well known.
For other values of $\beta$ the expression is more complicated.
For example for $\beta=4$ we find:
\be
\Omega^2=\frac{1}{1+\sin{2p}\sin{2q}}\; .
\ee 
Using the Hopital rule it can be {\it generally} proved that  $\Omega$ 
converges on the null boundaries of the cosmological diamond. 
It can also be generally proved that the only
divergence occurs for $\sin{2p}\sin{2q}=-1$ (i.e. for $-p=q=\pi/4$
and periodically related points), where the two null boundaries meet. 
Indeed away from the null boundaries 
of the cosmological patches $\Omega$ can only diverge if 
$\eta=i\frac{\pi}{\beta}$. This only has real solutions in 
$\{p,q\}$  for $\beta=2$. 
Thus the Penrose diagram (see Fig.~\ref{fig:AdS Penrose})
is the whole Einstein static Universe, if no
singularities isolate part of it. This is
true for $\beta>2$. Spatial infinity is at 
$\hat\chi=\pi/2$ and $\hat\xi=n\pi$. Unlike with AdS $\hat\chi=\pi/2$ is no longer ${\cal I}^\pm$.
As with AdS, it is impossible to conformally render finite time-like 
infinity without collapsing spatial distances to a point. 

\begin{figure}[h] 
\resizebox{0.5\columnwidth}{!}{\includegraphics{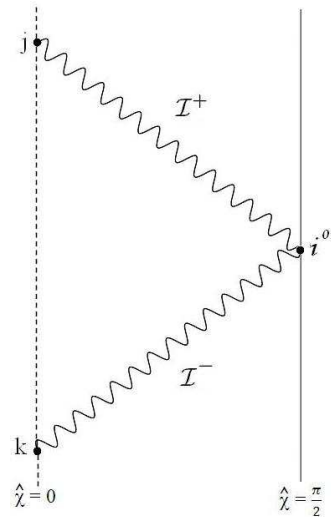}}
\caption{\label{fig:AdS Penrose2} The Penrose Diagram for our space-time
when $\beta<2$. }  
\end{figure}

However for $\beta<2$ there are singularities, associated with a 
diverging energy density. Homogeneity and isotropy preclude Weyl 
curvature, so all singularities must be Ricci singularities 
(since $w\neq 1/3$, a divergence of the Ricci tensor entails a
divergent  Ricci scalar). But the Ricci curvature 
is homogeneous in the foliations of constant $\xi$, and these 
leaves approach the null hypersurface that bounds the cosmological
half-diamond. Therefore the space is singular on these surfaces
(which are within a finite affine distance)
and the Penrose diagram of the space-time is that
depicted in Fig.~\ref{fig:AdS Penrose2}.

Finally note that a simple coordinate system can be obtained if the 
angular variables can be ignored. By defining a transformation as above 
but with  (\ref{tan1}) and (\ref{tan2}) replaced by
\bea
\tan p&=&{\rm tanh}\frac{\beta u}{4}\label{tan1b}\\
\tan q&=&{\rm tanh}\frac{\beta v}{4}\label{tan2b}
\eea
one obtains a metric of form
\be
ds^2=\frac{4\sec^{\frac{4}{\beta}}{\chi}}{\beta^2\cos^{1-\frac{2}{\beta}}(2p)
\cos^{1-\frac{2}{\beta}}(2q)}(-d{\hat\xi}^2+d{\hat\chi}^2+F(\hat\xi,\hat\chi)
d\Omega_2)
\ee
Conformal diagrams may be obtained with these coordinates but they 
hide a complex structure in the angular part of the metric 
$F(\hat\xi,\hat\chi)$ (with $F=\sin^2\hat\chi$ for $\beta=2$ only).

\section{Conclusions}
In summary we have used the set up of FRW cosmology as a springboard to
find simple generalizations of AdS space. They can be seen as $w\neq -1$
FRW solutions, but we extended them to their full manifolds. The embeddings
found reveal deformed hyperboloids with the same topology 
and causal structure as AdS. However the structure of Killing vector fields
is entirely modified and should be the subject of further enquiry.
We hope that these simple solutions might be useful in assessing the
generality of theorems proved for AdS, but conjectured to be true in 
an adapted form in more realistic spaces.  

To conclude we stress
that it wouldn't be difficult to generalize our constructions
to de Sitter-like space-times (for which all we'd need to do it change
the sign of the integration constant $C$ and the signature of the embedding
space). More spatial dimensions could also be trivial included.
Less trivial is the meaning of our construction when 
$w>-1/3$. Then the calculations can still be trivially carried out, but they lead to Euclidean metrics. These spaces were not
explicitly constructed here but may also be of interest.

\end{document}